\documentclass[sigconf]{acmart}

\usepackage{hyperref}
\usepackage{breakurl}
\usepackage{url}
\usepackage{balance}
\usepackage{booktabs} 

\usepackage{bbm}

\usepackage{amsmath}
\usepackage{amsthm}
\usepackage{enumitem}
\usepackage{amsfonts}

\usepackage{amssymb}
\usepackage{microtype}
\usepackage{tabularx}
\usepackage{siunitx}
\newtheorem{assumption}{Assumption}
 
\usepackage{multirow} 

\usepackage{graphicx} 

\usepackage[ruled]{algorithm2e} 

\setcopyright{rightsretained}

\usepackage{graphicx}
\usepackage{subcaption}

\AtBeginDocument{%
  \providecommand\BibTeX{{%
    \normalfont B\kern-0.5em{\scshape i\kern-0.25em b}\kern-0.8em\TeX}}}

\setcopyright{acmcopyright}
\copyrightyear{2021}
\acmYear{2021}
\acmDOI{10.1145/1122445.1122456}

\begin{document}
\fancyhead{}
\title{Computation Resource Allocation Solution in Recommender Systems}

\author{Xun Yang, Yunli Wang, Cheng Chen, Qing Tan, Chuan Yu, Jian Xu, Xiaoqiang Zhu}
 \affiliation{%
   \institution{Alibaba Group}
   \city{Beijing}
   \country{P.R.China}
 }
 \email{{vincent.yx,ruoyu.wyl,chencheng.cc,qing.tan,yuchuan.yc,xiyu.xj,xiaoqiang.zxq}@alibaba-inc.com}

\begin{abstract}
Recommender systems rely heavily on increasing computation resources to improve their business goal. By deploying computation-intensive models and algorithms, these systems are able to inference user interests and exhibit certain ads or commodities from the candidate set to maximize their business goals. However, such systems are facing two challenges in achieving their goals. On the one hand, facing massive online requests, computation-intensive models and algorithms are pushing their computation resources to the limit. On the other hand,  the response time of these systems is strictly limited to a short period, e.g. 300 milliseconds in our real system, which is also being exhausted by the increasingly complex models and algorithms.

How to efficiently utilize the computation resources and response time has become a surging problem for recommender systems. In this paper, we propose the computation resource allocation solution (CRAS) that maximizes the business goal with limited computation resources and response time. We comprehensively illustrate the problem and formulate such a problem as an optimization problem with multiple constraints, which could be broken down into independent sub-problems. To solve the sub-problems, we propose the revenue function to facilitate the theoretical analysis, and obtain the optimal computation resource allocation strategy. To address the applicability issues, we devise the feedback control system to help our strategy constantly adapt to the changing online environment. The effectiveness of our method is verified by extensive experiments based on the real dataset from Taobao.com. We also deploy our method in the display advertising system of Alibaba. The online results show that our computation resource allocation solution achieves significant business goal improvement without any increment of computation cost, which demonstrates the efficacy of our method in real industrial practice. 
\end{abstract}

%

\ccsdesc[500]{Information systems~Recommender systems}
\ccsdesc[300]{Information systems~Computational advertising}

\keywords{Recommender System, Computation Efficiency, Computational Advertising}

\maketitle

\section{Introduction}

	 A typical recommender system aims to maximize its business goal by exhibiting certain ads or commodities from the candidate set when a user visits the online site. Benefiting from extended user behavior data collected from online systems,  dedicated models are generally employed to capture user interests, which plays an important role in showing the optimal ads or commodities to maximize the business goal. Such models usually deliver excellent accuracy in estimating the user preferences, but are always accompanied with massive calculations. The common models include click-through rate (CTR) models \cite{zhou2018deep, chan2018convolutional, zhou2019deep} and conversion rate (CVR) models  \cite{yang2017bayesian}, which have been extensively studied in related areas.
	 
\begin{figure}
 	\centering    
    \includegraphics[width=0.48\textwidth]{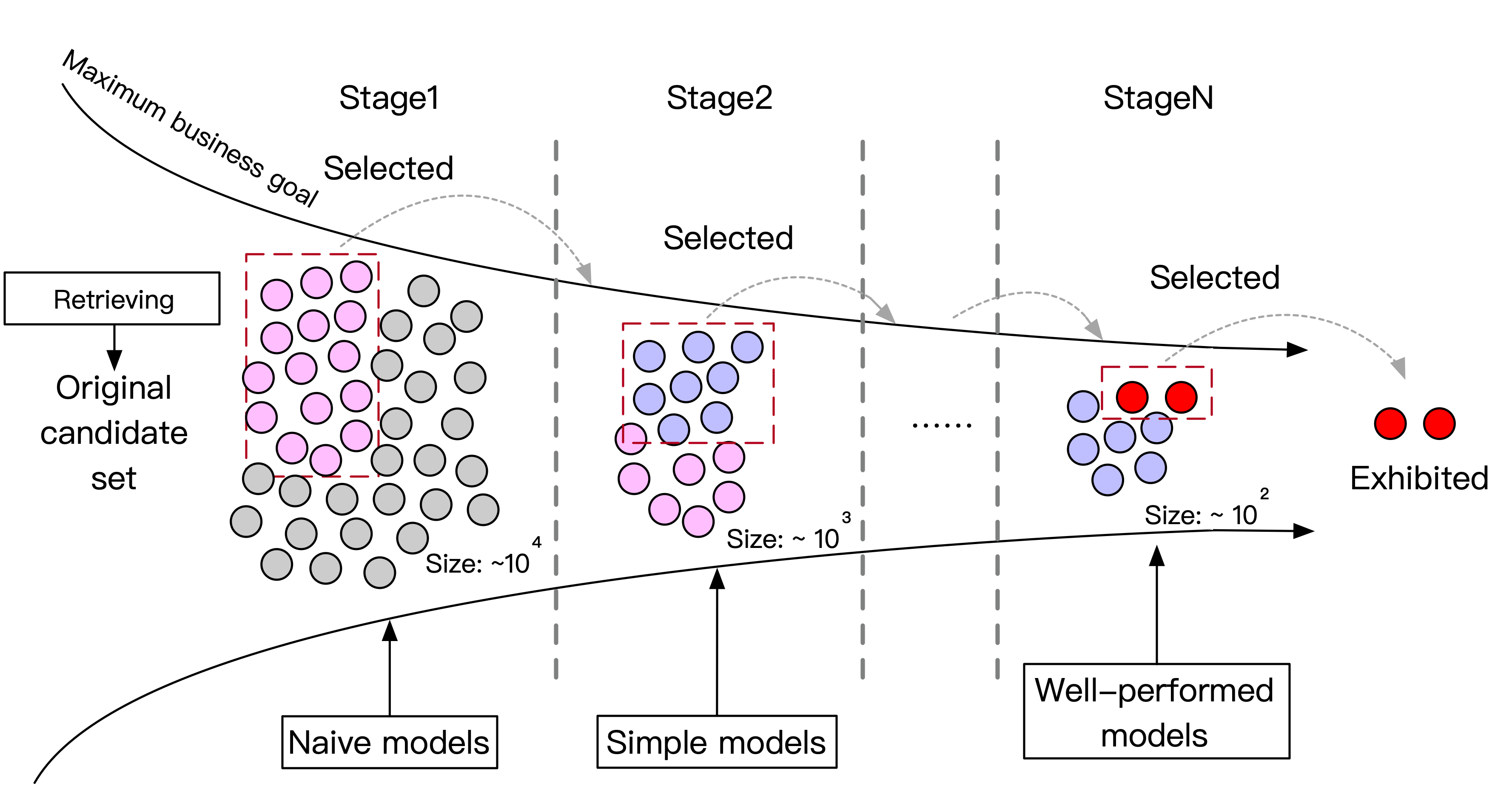}
    \caption{The general architecture of the recommender system. The system conducts successive stages of sorting and selecting based on the estimated business goal.}
    \label{fg:system_architect}
\end{figure}

	  However, well-performed models are hardly applied to the entire candidate set due to the large candidate set size. Taking the display advertising system of Alibaba  \footnote{Most advertising systems share the same architecture as recommender systems, so we refer to them all as recommender systems in this work.} for example, there are roughly 10 thousand candidate ads given an arbitrary user. Directly estimating the preference of one user on the whole candidate set with well-performed models would not only run out of the allowed response time, but also cost impractical computation resources. Therefore, industrial recommender systems commonly adopt a cascade architecture to trade-off among the computation cost, response time and model performance, which has been proved rather effective in practice \cite{covington2016deep}. The main idea of such architecture is to reduce the candidate set size by degrees using models of different complexity and computation requirement, which enables the system to enjoy well-performed models in a practical way. The general architecture of a recommender system is illustrated in Fig. \ref{fg:system_architect}. As shown in Fig. \ref{fg:system_architect}, the system conducts successive stages of sorting and selecting based on the estimated business goal with models of increasing estimation accuracy after retrieving the original candidate set. Take the display advertising system of Alibaba for example, whose business goal is to exhibit ads that maximize revenues. The whole process of this system after retrieving the original candidate set is divided into 3 successive stages: pre-ranking stage, coarse-ranking stage, and fine-ranking stage. Each stage sorts and selects ads from the current candidate set according to the estimated revenues, which is highly dependent on the CTR and CVR models. As ads are delivered to the next stage, the candidate set size becomes smaller and the model's estimation accuracy increases along with the computation cost. Such a design allows the system to take full advantage of the fruits of the state-of-the-art models while still being able to control the computation cost and response time.

	  One problem has never been addressed in such an architecture: how to balance the computation cost, response time, and model performance in the most efficient way. According to our empirical study, the candidate set size across each stage plays a crucial role in such a trade-off. Firstly, given models cost almost equal computation resources on every candidate ad, more candidate ads mean more computation cost. Also, more candidate ads that enjoy the models of high estimation accuracy result in more business goal improvement. That is to say, the candidate set size is the main factor that determines the computation cost and business goal. Secondly, more candidate ads result in more processing latency on sorting, network transmitting, and input/output operations, so the candidate set size largely determines the response time. Considering the above facts, we regard the candidate set size across stages as the key point in this work to analyze and optimize the system. However, today's systems, like the display advertising system of Alibaba, simply truncate the candidate set size by a fixed number in each stage based on the practical experience, and there are no guidelines on how to set the optimal candidate set size for each stage with limited computation resources and response time. Furthermore, most recommender systems truncate the candidate set size without regarding the specific user information. Considering the properties and features of each online request significantly differ from each other, this one-size-fits-all approach is not an efficient way to fully utilize the limited computation resources and response time. Although the recent work DCAF \cite{jiang2020dcaf} proposed a "personalized" computation resource allocation framework, this work focuses on one specific stage and the response time is not addressed. 
	  
	  In this work, we address the mentioned problem in the cascade architecture, and propose the computation resource allocation solution (CRAS) that maximizes the business goal of recommender systems given limited computation resources and response time. We give the full-covered solution including problem formulation, optimal computation resource allocation strategy, and industrial solution concerning the applicability issues.  
	  
	  In this paper, we formulate such a problem as an optimization problem with multiple constraints, which could be broken down into independent sub-problems. Solving the sub-problems, we propose the revenue function to facilitate the theoretical analysis. Given the revenue function, we obtain the optimal computation resource allocation strategy, whose meaning could be interpreted from the view of the economics. 
	  
	  In addition, a feedback control system is devised in this paper to address the changing online environment. Since the computation resource allocation strategy obtained based on the historical data may be non-optimal due to the changing distribution of online requests, we devise a feedback control system to deal with such an applicability issue. The feedback control system could constantly adjust the computation resource allocation strategy around the optimal without increasing system burdens while facing online traffic changes.
	  
	  Moreover, the proposed computation resource allocation solution is implemented and evaluated on real industrial datasets, and has been deployed in the display advertising system of Alibaba.  Experiments on the real datasets and online results show that our computation resource allocation solution achieves better business goal without increasing any computation cost compared with the baseline, which demonstrates the effectiveness of our method. The main contributions of our work can be summarized as follows:  
	  
\begin{enumerate}[align=right,leftmargin=0.17in]
\item We propose the computation resource allocation solution that maximizes the business goal of recommender systems with limited computation resources and response time. We formulate the problem as an optimization problem with multiple constraints, and derive the optimal computation resource allocation strategy.  
\item A feedback control system is devised to address the applicability issue when applying the computation resource allocation strategy in the industrial environment.
\item Extensive experiments and online results demonstrate the effectiveness of our method. 
\end{enumerate}

\section{Methodology}

In this section, we describe the problem in length and formulate it as an optimization problem with multiple constraints, and then break down the complex original problem into independent sub-problems. We derive the optimal solution of the sub-problems by leveraging the primal-dual method, and interpret its meaning from the view of economics. In the following discussion, we base our discussion on the display advertising system of Alibaba to facilitate the narrative, and any technique involved in this work could be easily generalized to other cascade-architecture systems. 

\subsection{Problem Formulation}

\begin{figure}
 	\centering    
    \includegraphics[width=0.48\textwidth]{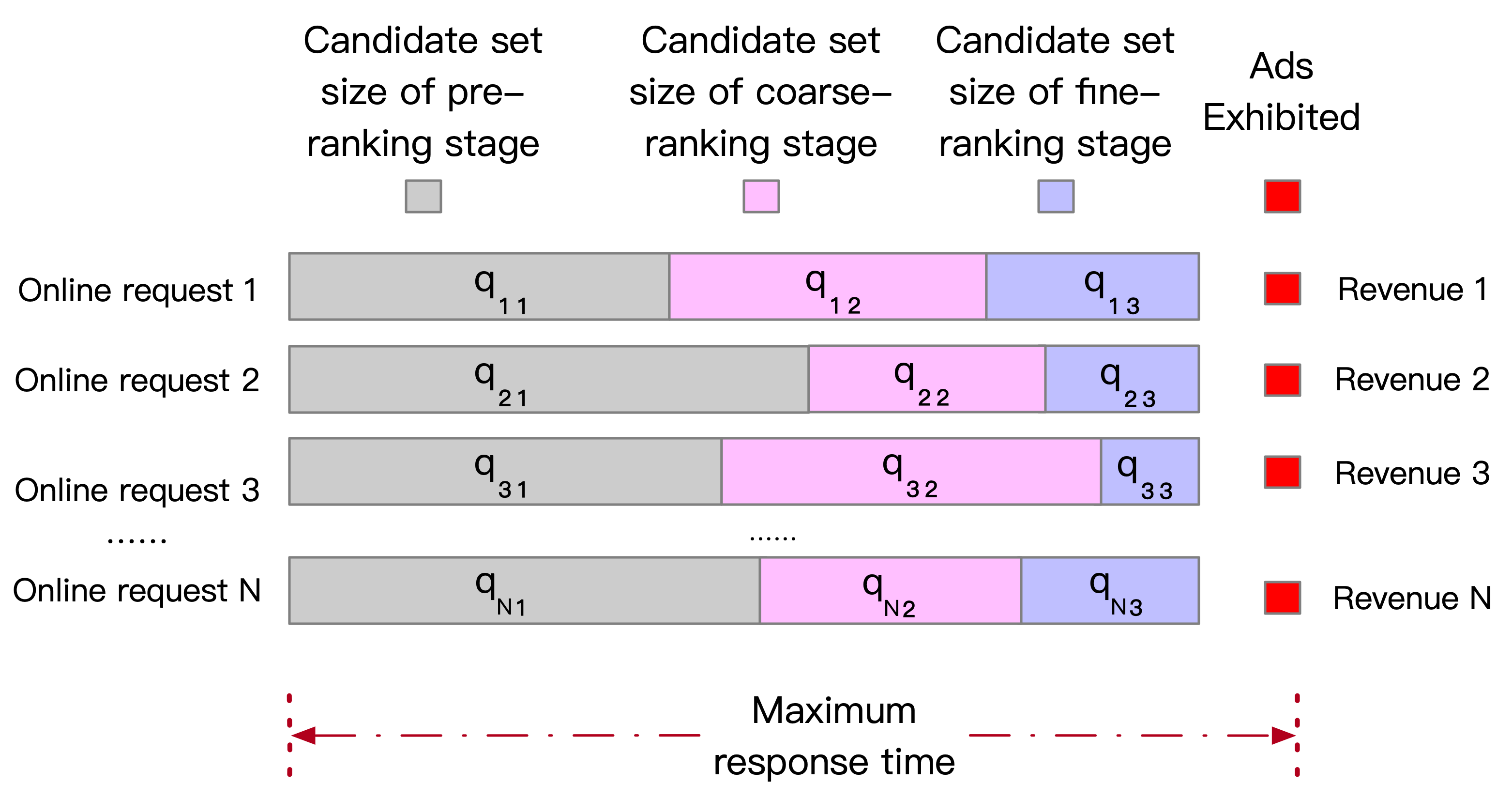}
    \caption{A graphical illustration of the problem. The key problem is to find the optimal candidate set size of each stage for each online request under the constraint of computation cost and response time.}
    \label{fg:allocation_problem}
\end{figure}

As we discussed before, the candidate set size across stages is the main factor that determines computation cost, response time, and business goal. Therefore, the key problem is to find the optimal candidate set size of each stage for each online request under the constraint of computation cost and response time. As illustrated in Fig. \ref{fg:allocation_problem}, we suppose there are $N$ online requests in a time session. We use $q_{*1}$, $q_{*2}$, $q_{*3}$ to represent the allocated candidate set size for pre-ranking stage, coarse-ranking stage and fine-ranking stage respectively, and $q_{i*}$ to represent the setting of the specific online request $pv_i$. For example, $q_{i2}$ represents the allocated candidate set size of coarse-ranking stage for the online request $pv_i$. Online request $pv_i$ would yield the revenue $revenue_i$ given the candidate set size of each stage (i.e. $q_{i1}$, $q_{i2}$, $q_{i3}$). We use the joint revenue function $Y(pv_i, q_{i1}, q_{i2}, q_{i3})$ to represent $revenue_i$. We further assume that the revenue function $Y(pv_i, q_{i1}, q_{i2}, q_{i3})$ is multiplicatively separable and get the reformed function in Eq. \eqref{yield_function1}, where $M_i$ is the maximum revenue we could ideally achieve without any truncating. $Y_1$, $Y_2$, and $Y_3$ are all no greater than $1.0$ and represent the discounter caused by the truncating process of each stage respectively, which aligns with the problem in our successive truncating setting and helps to avoid unnecessary complexities. 

\begin{align}
			&revenue_i & &= Y(pv_i, q_{i1}, q_{i2}, q_{i3}) \\ 
			& & &=  M_i \cdot Y_1(pv_i, q_{i1}) \cdot Y_2(pv_i, q_{i2}) \cdot Y_3(pv_i, q_{i3}) &\label{yield_function1}
\end{align}

As the computation cost of each stage increases linearly against its candidate set size, we use the sum of the allocated candidate set size to represent the total computation cost. It is worth noting that the models across different stages are usually deployed in independent computation servers with heterogeneous computation resources \cite{wang2020cold, pi2019practice}, so we set the computation cost constraints in each stage as independent constraints. We use $C_1$, $C_2$, and $C_3$ to represent the computation cost constraints in each stage respectively. Meanwhile, we use $L(q_{i1}, q_{i2}, q_{i3})$ to represent the response time given the candidate set size of each stage, and the maximum response time is $T$. It is worth mentioning that the response time only depends on the candidate set size, and the information of online request $pv_i$ does not affect it. To sum up, we aim to maximize the total revenues (business goal) with the constraints of computation cost and response time, and formulate the computation resource allocation problem as shown in \eqref{lp0}.

 One challenging part in the problem \eqref{lp0} is the fact that it is hard, even impractical to obtain the response time function $L(q_{i1}, q_{i2}, q_{i3})$. In our theoretical assumption, the total response time only depends on $q_{i1}$, $q_{i2}$ and $q_{i3}$ for online request $pv_i$. However, it is vulnerable to many factors in an industrial online environment such as network transmission, machine load, and hardware performance, which is unpredictable and changeable. Facing the inconsistency of the complex industrial online environment, it is reasonable and necessary to sacrifice certain theoretical optimality in exchange for industrial applicability and system robustness. Therefore, we slightly strengthen the constraint InEq. \eqref{rt_constraint} and modify it as stated in InEq. \eqref{rt_constraint1}. Since the response time is monotonously increasing against the candidate set size of each stage, satisfying the constraint InEq. \eqref{rt_constraint1} would naturally make sure that the constraint InEq. \eqref{rt_constraint} is achieved in theory. Such a modification brings us two benefits: 1) it allows our analysis to be practical while facing inconsistent online systems. Theoretically modeling the function $L(q_{i1}, q_{i2}, q_{i3})$ is not only unreliable and impractical, but also brings difficulty in applying the method in different systems. Setting $D_1$, $D_2$ and $D_3$ as independent constraints helps to simplify the analysis and make the analysis more generalized across different systems. 2) independent response time restrictions help to improve the robustness of the system. Such a setting makes the system more controllable, and humans can intervene immediately when there is a problem. For example, $D_1$, $D_2$, and $D_3$ could be rapidly adjusted and improved according to the system monitors when there is a great change in the online system. 
 
 \begin{align}
			 &\underset{q_{i1}, q_{i2}, q_{i3}}{\textup{max}} & & \displaystyle\sum\limits_{i=1...N}M_i \cdot Y_1(pv_i, q_{i1}) \cdot Y_2(pv_i, q_{i2}) \cdot Y_3(pv_i, q_{i3}) \tag{P0}\label{lp0} \\
			&\textup{s.t.} & &  \displaystyle\sum\limits_{i=1...N}q_{i1} \leq C_1  \\
			& & &  \displaystyle\sum\limits_{i=1...N}q_{i2} \leq C_2  \\
			& & &  \displaystyle\sum\limits_{i=1...N}q_{i3} \leq C_3  \\
		&	& &  L(q_{i1}, q_{i2}, q_{i3}) \leq T, \forall i   \label{rt_constraint}\\
		& & &  q_{i1}, q_{i2}, q_{i3}  \geq 0, \forall i \nonumber 
\end{align}

As we replace the constraint InEq. \eqref{rt_constraint}
 with the constraint InEq. \eqref{rt_constraint1}, the original problem \eqref{lp0} could be broken down into three independent sub-problems since there is no joint interaction among $q_{i1}$, $q_{i2}$ and $q_{i3}$. Therefore, we could independently solve the sub-problems to obtain the global optimal solution, and all sub-problems share the same formulation as stated in \eqref{lp1} \footnote{We slightly abbreviate the subscript to avoid redundancy. Taking fine-ranking for example, $q_{i3}$ is abbreviated as $q_i$, $C_3$ is abbreviated as $C$, and $D_3$ is abbreviated as $D$.}, where $Y(q_i, pv_i)$ could be regarded as the new revenue function that represents the achieved revenue given $q_i$ and $pv_i$ in this specific stage without any other truncating stage. Taking the fine-ranking for example, we could obtain the sub-problem \eqref{lp1} by setting $Y_1$ and $Y_2$ constantly equal to $1.0$ (or any other constant value) and combining $M_i$ and $Y_3$ into the new revenue function $Y(q_i, pv_i)$. It is worth noting that it does not affect the optimal solution in the sub-problem \eqref{lp1} by setting the discounter of other stages constantly equal to $1.0$ since the contribution for the revenue of each stage is independent. In our following discussion, we focus on solving the sub-problem of the fine-ranking stage without loss of generality. 
 
 \begin{align}
&q_{i1} \le D_1,\;
q_{i2} \le D_2, \;
q_{i3} \le D_3 \label{rt_constraint1} \\
&\text{where  } L(D_1, D_2, D_3) \le T  \nonumber 
\end{align}

\begin{align}
			 &\underset{q_i}{\textup{max}} & & \displaystyle\sum\limits_{i=1...N}Y(q_i, pv_i) \tag{P1}\label{lp1} \\
			&\textup{s.t.} & &  \displaystyle\sum\limits_{i=1...N}q_i \leq C \nonumber \\
		&	& &  q_i \leq D, \forall i  \nonumber  \\
		& & &  q_i \geq 0, \forall i \nonumber 
\end{align}

\subsection{Revenue Function}

The problem \eqref{lp1} is an optimization problem with linear constraints. The key challenge is that $Y(q_i, pv_i)$ is unknown. Before we obtain the general form of function $Y(q_i, pv_i)$, we assume $Y(q_i, pv_i)$ should have the following two properties in general: 
\begin{assumption}\label{as_1}
$Y(q_i, pv_i)$ is monotonously increasing with respect to $q_i$.
\end{assumption}

\begin{assumption}\label{as_2}
$ \frac{dY(q_i, pv_i)}{dq_i}$ is monotonously decreasing with respect to $q_i$. 
\end{assumption}

Assumption \ref{as_1} is straightforward. When $q_i$ increases, more ads are sent to the fine-ranking stage and enjoy complex and expressive models, which should lead to an increment of revenue. Assumption  \ref{as_2} actually describes the general situation in real-world and points out that the marginal utility of the system should decrease while investing more computation resources. The decreasing marginal utility phenomenon described in Assumption  \ref{as_2} is rather common in many applications \cite{al2005optimal, lehmann2006combinatorial} and is reasonable in the online advertising and recommendation scenarios \cite{wang2011utilizing}.

\begin{figure}
    \centering
    \begin{subfigure}[b]{0.2\textwidth}
        \includegraphics[width=\textwidth]{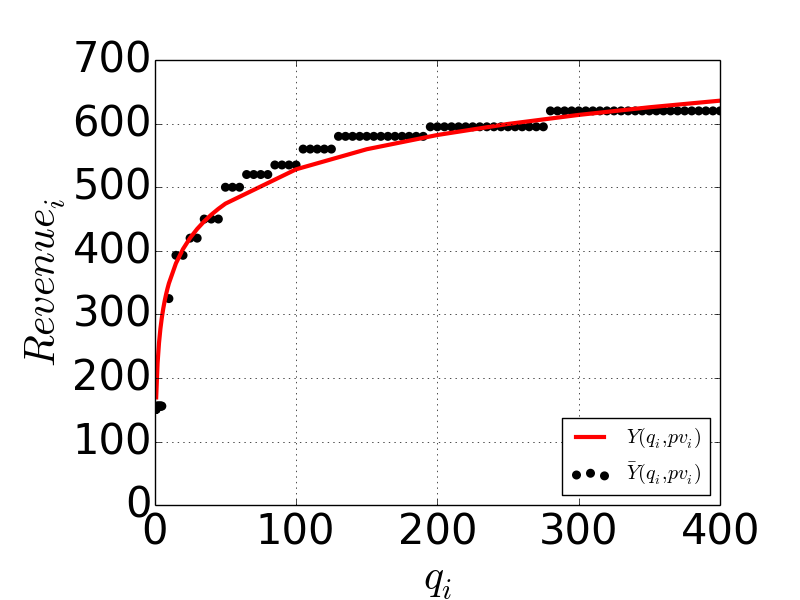}
        \caption{Example 1}

        \label{fg:revenue_function_fig1}
    \end{subfigure}
    \begin{subfigure}[b]{0.2\textwidth}
        \includegraphics[width=\textwidth]{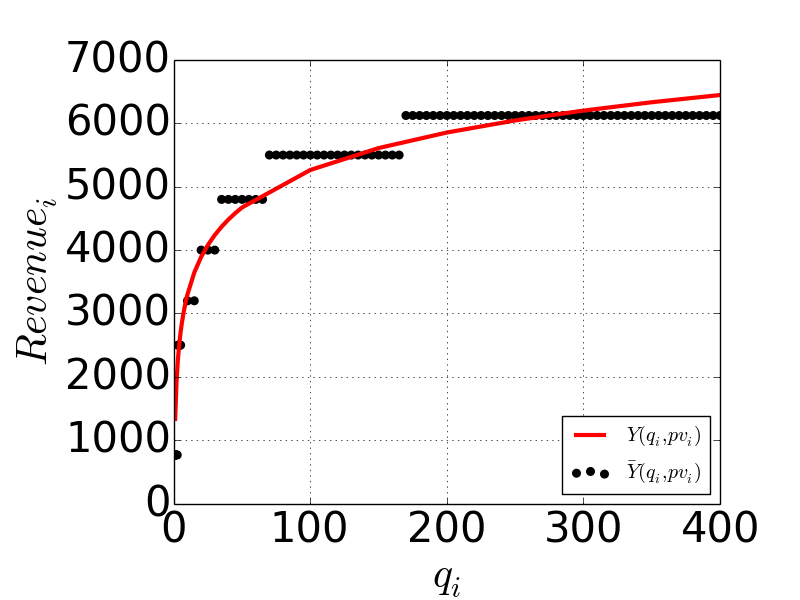}
        \caption{Example 2}
        \label{fg:revenue_function_fig2}
    \end{subfigure}
    \caption{The revenue functions of two example online requests in the fine-ranking stage by offline simulations. The revenue function could be fitted by a natural logarithm function with neglectable deviation.}\label{fg:revenue_function_fig}
\end{figure}

We could obtain the revenue function by offline simulations. The data of the whole ad-selecting procedure in most online systems are logged and dumped, so that we are able to calculate the revenue for every online request $pv_i$ with arbitrary $q_i$ by offline simulations. We use $\bar{Y}(q_i, pv_i)$ to represent the original revenue function obtained by offline simulations. The revenue function $\bar{Y}(q_i, pv_i)$ of two example online requests based on the real data is illustrated in Fig. \ref{fg:revenue_function_fig}. It is worth noting that the revenue function $\bar{Y}(q_i, pv_i)$ is a discrete function since the candidate set size is an integer. Also, it is a step-like function because a small change of $q_i$ may not influence the revenue in practice. Directly applying such a function in the problem \eqref{lp1} brings us unnecessary complexity and difficulty. Therefore, we propose to replace the original revenue function with well-defined functions to facilitate the analysis, which incurs little influence on the optimal solution as we show in the following experiments. Specifically, we adopt the natural logarithm ($Ln$) functions \footnote{We also tried polynomial functions and square root functions, and logarithm functions deliver the best performance in both theoretical analysis and industrial practice.} to fit the revenue function achieved by offline simulations due to the following two reasons: 1) $Ln$ functions naturally align with the above two assumptions; 2) $Ln$ functions are of simple formulation that could largely facilitate the theoretical analysis with trivial deviation from the original revenue function, which is demonstrated in the Fig. \ref{fg:revenue_function_fig}. Therefore, we design the revenue function as Eq. \eqref{revenue_function} states, where $R_i$ and $B_i$ are hyperparameters of $pv_i$ that determine its revenue function.

\begin{equation}\label{revenue_function}
Y(q_i, pv_i) = R_i \cdot Ln q_i + B_i
\end{equation}

\subsection{Optimal Allocation Strategy}

Given the revenue function $Y(q_i, pv_i)$, we restate \eqref{lp1} as \eqref{lp2}.

\begin{align}
			 &\underset{q_i}{\textup{max}} & & \displaystyle\sum\limits_{i=1...N}R_i \cdot Lnq_i +B_i \tag{P2}\label{lp2} \\
			&\textup{s.t.} & &  \displaystyle\sum\limits_{i=1...N}q_i \leq C \label{con_C} \\
		&	& &  q_i \leq D, \forall i \label{con_D} \\
		&	& &  q_i \geq 0, \forall i \label{con_E}
\end{align}

The problem \eqref{lp2} is a convex optimization problem, which could be regarded as a primal problem. According to the primal-dual theory \cite{boyd2004convex}, a primal problem could be converted to a dual problem, and the optimal solution would remain the same as long as the strong duality holds \cite{slater2014lagrange}, which is applicable in our case. The dual problem is stated formally in \eqref{lp3}, where the new objective function is abbreviated as $Dual$ and it does not influence our following demonstration. It is worth noting that $\alpha$, $\beta_i$ and $\gamma_i$ are Lagrange  Multipliers respectively introduced by constraints InEq. \eqref{con_C}, InEq. \eqref{con_D} and InEq. \eqref{con_E}.

\begin{align}
			 &\underset{\alpha, \beta_i, \gamma_i}{\textup{min}} & & Dual(\alpha, \beta_i, \gamma_i) \tag{P3}\label{lp3} \\
			&\textup{s.t.} & &  q_i(\alpha +\beta_i - \gamma_i) = R_i \label{res_q} \\
		&	& &  \alpha \geq 0 \nonumber \\
		&	& &  \beta_i, \gamma_i \geq 0, \forall i \nonumber 
\end{align}

According to the primal-dual theory, the constraint Eq. \eqref{res_q} in \eqref{lp3} must hold if the optimal solution is achieved, so we could firstly derive the optimal $q_i ^*$ by solving Eq. \eqref{res_q} with representation of $\alpha$, $\beta_i$ and $\gamma_i$. Therefore, we obtain the optimal solution $q_i ^*$ as shown in Eq. \eqref{optimal_q}, where $\alpha ^*$, $\beta_i ^*$ and $\gamma_i^*$ are the optimal value in the corresponding dual problem.  It needs to be noted that Eq. \eqref{optimal_q} does not explicitly tell the value of $\alpha^*$, $\beta_i^*$ and $\gamma_i^*$, which could be obtained by developed programming algorithms. In the following discussion, we shall introduce an effective way to directly obtain the optimal $\alpha^*$, $\beta_i^*$ and $\gamma_i^*$ without unnecessary mathematical calculations.

\begin{equation} \label{optimal_q}
\begin{aligned}
q_i^* = \frac{R_i}{\alpha^* + \beta_i^* - \gamma_i^*} \ \ \\
\end{aligned}
\end{equation}

Please recall that $\beta_i$ and $\gamma_i$ are Lagrange Multipliers introduced respectively by constraints
InEq. \eqref{con_D} and InEq. \eqref{con_E}. According to the theorem of complementary slackness \cite{boyd2004convex}, Eq. \eqref{slackness_d} and Eq. \eqref{slackness_e} could be derived, and we have the following two statements:1) $\beta_i$ equals $0$ if $q_i$ is less than $D$; 2) $\gamma_i$ equals $0$ if $q_i$ is greater than $0$. In other words, $\beta_i$ and $\gamma_i$ are both zero as long as $q_i$ lies in the interval of $(0, D)$. Therefore, we could reform Eq. \eqref{optimal_q} and obtain our optimal computation resource allocation strategy in Eq. \eqref{optimal_q_new}, where $q_i$ is truncated by $D$ if $R_i/\alpha^*$ is greater than $D$. 

\begin{equation} \label{slackness_d}
\begin{aligned}
\beta_i^* \cdot (q_i^* -  D) = 0 \ \ \\
\end{aligned}
\end{equation}

\begin{equation} \label{slackness_e}
\begin{aligned}
\gamma_i^* \cdot q_i^* = 0 \ \ \\
\end{aligned}
\end{equation}

\begin{equation} \label{optimal_q_new}
\begin{aligned}
q_i^* = \frac{R_i}{\alpha^*}, \;\;\;\;\;  0 < q_i^* \le D
\end{aligned}
\end{equation}

Having derived the optimal computation resource allocation strategy, we take a discussion on its intrinsic meanings. The most prominent property of the strategy is that the revenue function's derivative with respect to $q_i$ is the same across all online requests with $q_i^*$ between $0$ and $D$. We demonstrate it in Eq. \eqref{equilibrium}, which reveals the fact that the marginal utility of every online request is equal to $\alpha^*$ with the optimal strategy. From the view of economics, it means that the system has reached an equilibrium point that any transfer of computation resources among online requests would no longer increase the total revenue. We could regard $\alpha^*$ as the current marginal utility of the whole system as we invest more computation resources.

\begin{equation} \label{equilibrium}
\begin{aligned}
\frac{dR_i Lnq_i+B_i}{d q_i}|_{q_i = q_i^*} = \frac{R_i}{q_i} |_{q_i = q_i^*} = \alpha^*, \forall i
\end{aligned}
\end{equation}

\section{System Design}

In this section, we put the proposed computation resource allocation strategy into practice. We first illustrate the overview of the online system, and then address the applicability issues as we apply the computation resource allocation strategy in the industrial scenario. Finally, we present our feedback control system to deal with such issues.

\subsection{System Overview}
 
\begin{figure}
 	\centering    
    \includegraphics[width=0.48\textwidth]{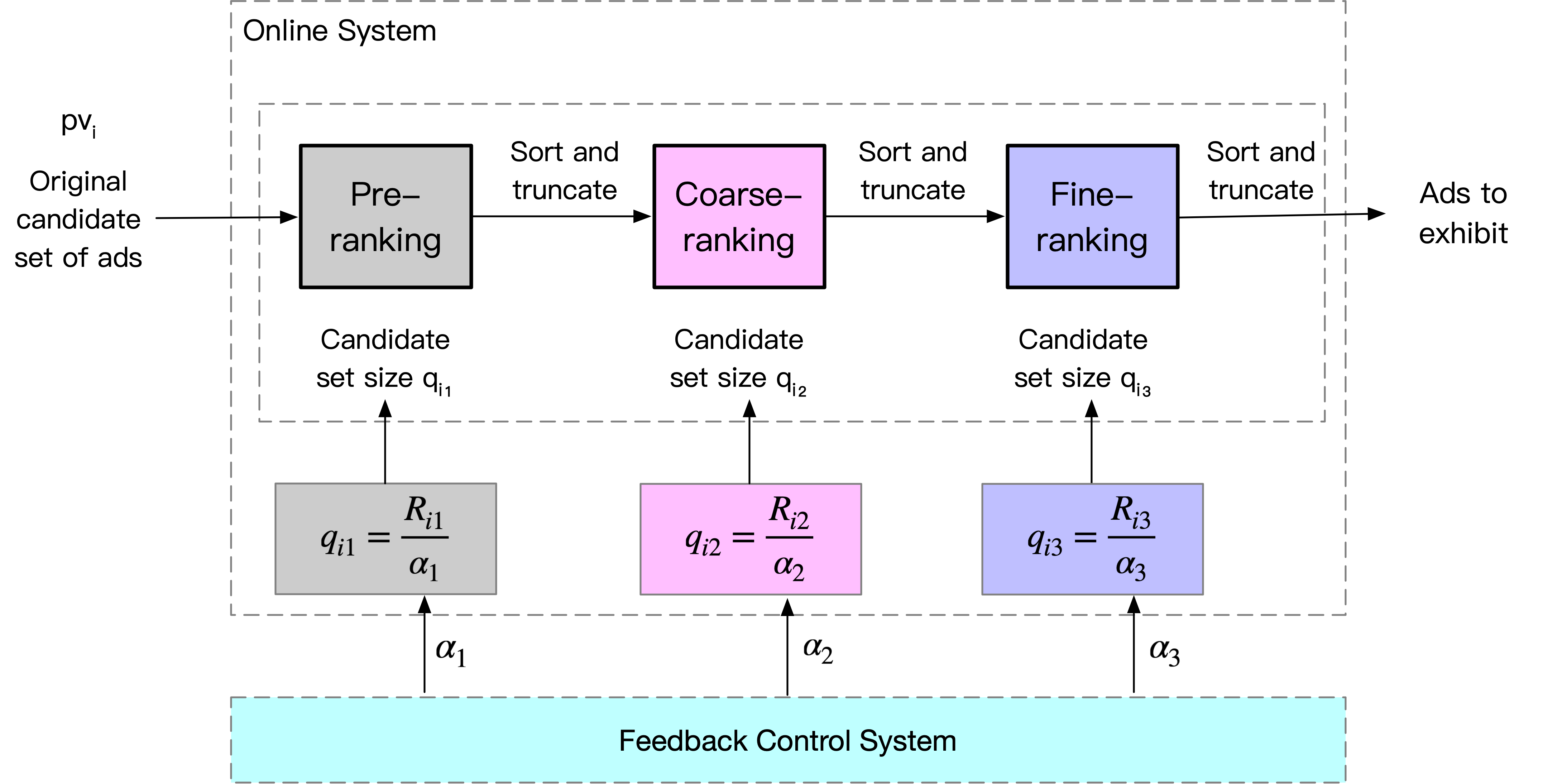}
    \caption{The overview of the online system. The candidate set size of each stage is independently determined by the computation resource allocation strategy.}
    \label{fg:system_overview}
\end{figure}

We illustrate the overview of the online system in Fig. \ref{fg:system_overview}. As an online request is triggered by the user, the original candidate set of ads would successively go through the pre-ranking stage, coarse-ranking stage and fine-ranking stage. The candidate set size of each stage $q_{i1}$, $q_{i2}$ and $q_{i3}$ are independently determined by the computation resource allocation strategy of each stage, which is introduced in the latest section. Specifically, taking the fine-ranking stage for example, the candidate set size $q_{i3}$ is calculated by $R_{i3}$ and $\alpha_3$ \footnote{We slightly abuse the subscript without incurring confusion.}, where $q_{i3}$ is no greater than $D_3$ to assure the response time constraint. In addition, a feedback control system is deployed to assist the online system to assure the computation cost constraint by dynamically adjusting $\alpha_3$, which would be introduced in the following sections. To facilitate the narrative, we base our discussion on the fine-ranking stage, and omit the unnecessary subscript when we address the specific stage. For example, while we are addressing the fine-ranking stage, we replace $R_{i3}$ with $R_i$ to avoid redundancy. 

\subsection{Applicability Issue}

One challenge we are facing is that we could not obtain the revenue function, i.e. $R_i$ of the online request beforehand in the online environment. In our previous discussion, we obtain the revenue function by offline simulations based on the logged data and fit it by $Ln$ functions, which is afterward the event and prohibited in the online environment. In practice, the revenue function needs to be obtained before the online request comes. Actually, the property of the online request is mainly determined by the visiting user itself. The feature and characteristics of online requests differ from each other because they are triggered by different online users. It is natural and reasonable to assume that multiple online requests triggered by the same online user would deliver the same properties, which means such online requests share the same revenue function. Therefore, we could obtain the revenue function of every online user in advance, and fetch the revenue function online when the corresponding user triggers the online request.

Another challenge is that the optimal value of $\alpha$ (i.e. $\alpha ^*$) is hard to be derived in the online environment. As shown in \eqref{lp2}, in order to solve the optimization problem and obtain the optimal hyperparameter $\alpha^*$ in Eq. \eqref{optimal_q_new}, we need to know exactly the information of every online request.  Apart from the fact that it is hard, even not practicable, to access such information beforehand, the changing online environment makes it difficult to predict. One straightforward way is to obtain the optimal value based on the historical data and apply it to the current time session. However, one strong assumption made in such a method is that the distribution of the online requests is stationary, which is rather rare in real-world online applications. Therefore, we propose our feedback control system to solve such a problem in the next section.

\subsection{Feedback Control System}

As discussed in the latest section, $\alpha^*$ is hard to be derived beforehand for the current time session. In addition, given the dynamic online environment, $\alpha^*$ obtained based on the historical time sessions could be non-optimal. Therefore, we propose to constantly adjust the $\alpha$ to approach the ideal $\alpha^*$ across time sessions.

To address the above issue, we revisit the optimal computation resource allocation strategy in Eq. \eqref{optimal_q_new}, where $\alpha$ is introduced from the dual space by the constraint InEq. \eqref{con_C}. Considering the fact that the revenue is maximized only if the equality holds in InEq. \eqref{con_C} (or otherwise $\alpha^*$ is zero, which makes no sense in our situation), $\alpha^*$ would ensure that the sum of candidate set size (i.e. computation cost) equals $C$. Furthermore, it is obvious that the computation cost is monotonically decreasing with respect to $\alpha$.  In other words, any $\alpha$ corresponds to an optimal computation resource allocation strategy with the corresponding computation cost constraint. Therefore, we could simply set the sum of $q_i$ equal to $C$ by adjusting $\alpha$, and thus the current $\alpha$ is guaranteed to be optimal. To sum up, we claim that we could simply adjust $\alpha$ to regulate the sum of $q_i$ around $C$, and thus make sure the $\alpha$ is around $\alpha^*$.  By doing so, we transform such an applicability issue into a feedback control problem.

\begin{figure}
 	\centering    
    \includegraphics[width=0.48\textwidth]{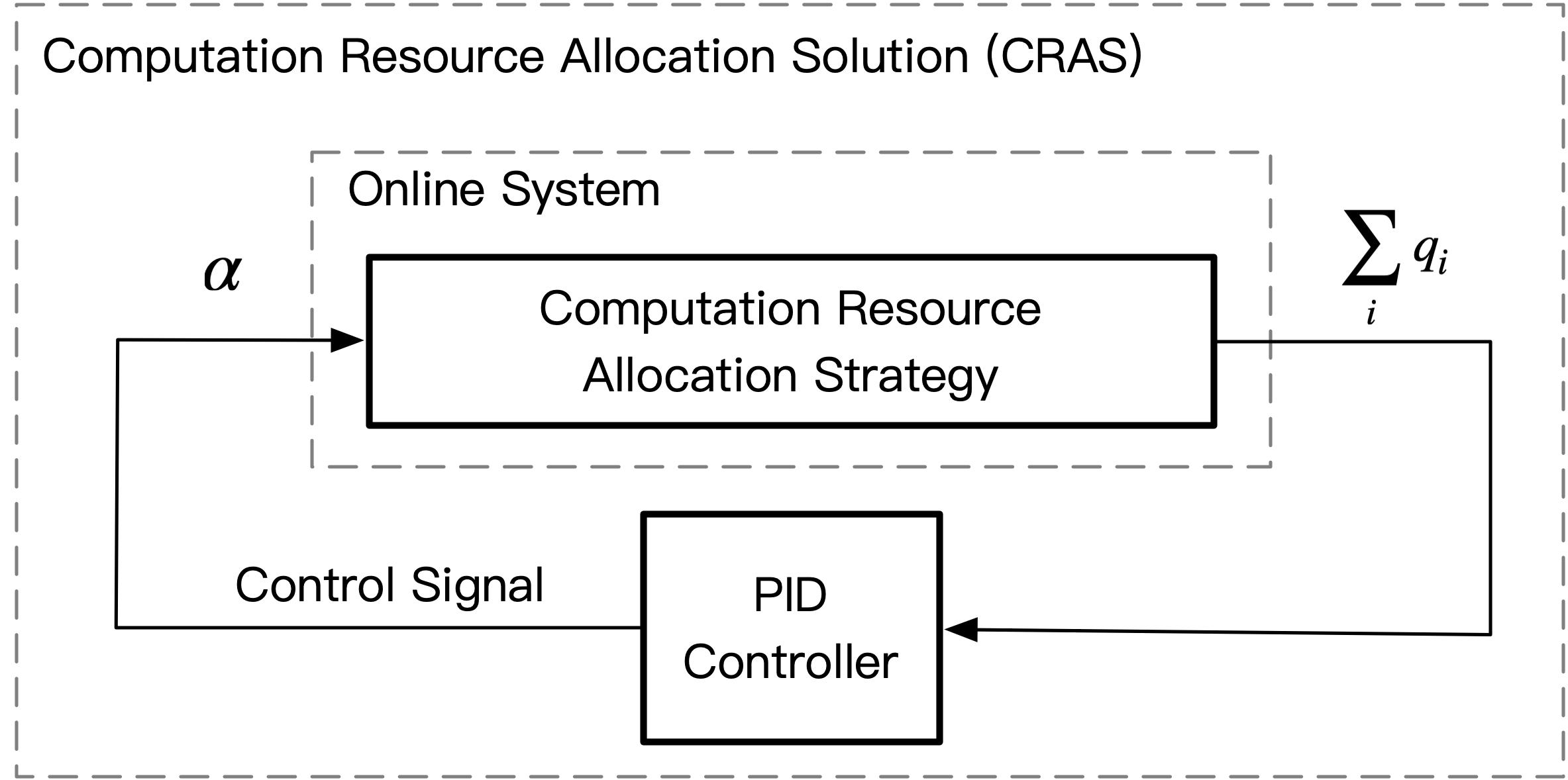}
    \caption{Computation resource allocation solution, where $\alpha$ is constantly adjusted to approach $\alpha^*$.}
    \label{fg:pid_control}
\end{figure}

 Proportional-Integral-Derivative (PID) controller \cite{bennett1993development} is the most widely adopted feedback controller in the industry. It is known that a PID controller delivers the best performance in the absence of knowledge of the underlying process with prominent robustness. A PID controller continuously calculates the error $e(t)$ between the measured value $y(t)$ and the reference $r(t)$ at every time step $t$, and produce the control signal $u(t)$ based on the combination of proportional, integral, and derivative terms of $e(t)$. The control signal $u(t)$ is then sent to adjust the system input $x(t)$ by the actuator model $\phi(x(0), u(t))$. It is practical and common to use discrete time step ($t_1, t_2, ...$) in online advertising and recommendation scenario, so the process of PID could be formulated as following equations, where $k_p$, $k_i$, and $k_d$ are the weight parameters of a PID controller. We list the specific formulations in Eq. \eqref{pid_et}, Eq. \eqref{pid_ut} and Eq. \eqref{pid_xt}. To sum up, we design the computation allocation resource solution (CRAS) with the feedback control system as illustrated in Fig. \ref{fg:pid_control}, where the feedback control system is independently deployed in each stage to adjust the corresponding $\alpha$.

\begin{equation} \label{pid_et}
\begin{aligned}
e(t) = r(t) - y(t) \ \ \\
\end{aligned}
\end{equation}	
	
\begin{equation} \label{pid_ut}
\begin{aligned}
u(t) = k_p  e(t) + k_i  \sum\limits_{i=1...t}e(k) + k_d  (e(t) - e(t-1)) \ \ \\
\end{aligned}
\end{equation}

\begin{equation} \label{pid_xt}
\begin{aligned}
x(t+1) = \phi (x(0), u(t)) \ \ \\
\end{aligned}
\end{equation}

\section{Empirical Study}
	
In this section, we conduct comprehensive experiments to demonstrate the effectiveness of our method. Following a detailed description of the system setting, dataset and evaluation metrics, we illustrate our implementation details in length. Experiments are conducted on the real-world dataset to evaluate the proposed computation resource allocation solution. Also, we deploy our method in the display advertising system of Alibaba to evaluate its effectiveness in industrial practice.

\subsection{Experiment Setup} 
\subsubsection{System Description}

The business goal of the display advertising system of Alibaba is to exhibit ads that maximize revenues. The whole process of this system could be divided into 3 successive stages: pre-ranking stage, coarse-ranking stage and fine-ranking stage. Each stage sorts and selects ads from the current candidate set according to the estimated revenues, which is highly dependent on the CTR and CVR models. As ads are delivered to the next stage, the candidate set size becomes smaller and the model's estimation accuracy increases along with the computation cost. Specifically, in the pre-ranking stage, CTR and CVR models are statistical models, which are rather simple and capture only the history information of the ad. In the coarse-ranking stage, the models adopt the light deep neural network architecture\cite{yi2019sampling}, which captures the user information and ad information in an efficient way. In the fine-ranking stage, the models are deep neural network models with complex and deep structures \cite{zhou2019deep}, which significantly increases the estimation accuracy as well as the computation cost.

\subsubsection{Dataset} 

The display advertising system of Alibaba could log the detailed information throughout the online process, so we construct the dataset based on the online logs. We sample millions of online requests as well as their information on Taobao.com. Each online request contains the information of the user and all candidate ads, which is required by the CTR and CVR models to estimate the revenue. Given user information, ad information and context information, the estimated revenue could be re-produced in the offline environment with corresponding CTR and CVR models across the pre-ranking stage, coarse-ranking stage and fine-ranking stage.

\subsubsection{Metrics} 

The main metrics we concern about in the recommender system are the revenue and computation cost. The total revenue achieved is a straightforward metric to evaluate the performance of the system since it is the business goal that we are maximizing. It is worth mentioning that the revenue is zero if the response time exceeds its limit, so the revenue could naturally reflect the general status of achieving the response time constraint. Since the computation cost is linear against the candidate set size in each stage, we use the sum of the candidate set size of all online requests to quantify the computation cost in each stage. As for the performance of the feedback control system, we graphically illustrate the environment changes and the system adjustment to evaluate the control capability.

\subsection{Implementation Details} 

\subsubsection{Revenue Function Fitting}
As illustrated in Fig. \ref{fg:revenue_function_fig}, we propose to replace the original revenue function, which is obtained by the offline simulation, with the logarithm function to facilitate the theoretical analysis. Such approximation incurs trivial influence as we will demonstrate in the following experiment. In this section, we describe our method to obtain the logarithm function. To facilitate the narrative, we assume the original revenue function achieved by the offline simulation is $\bar{Y}(q_i , pv_i)$, and the logarithm function is $Y(q_i, pv_i)$, whose formulation is stated in Eq. \eqref{revenue_function}. Our aim is to find the optimal $Y(q_i, pv_i)$ that is the most similar to the $\bar{Y}(q_i, pv_i)$ by minimizing the Mean Squared Error (MSE) between them, which is formed in the problem \eqref{eq:mse}. It is worth noting that we adopt the mean squared error to quantify the similarity between $Y$ and $\bar{Y}$, and one may also adopt other metrics such as absolute error, which does not make a big difference in our situation since the similarity is good enough as we will show in the following experiment. In our implementation, we leverage the well-developed algorithms in Scipy\footnote{\url{https://www.scipy.org/}} to derive the hyperparameters $R_i$ and $B_i$ of $Y(q_i, pv_i)$ by solving the problem \eqref{eq:mse}.

\begin{align}
			 &\underset{R_i, B_i}{\textup{argmin}} & & \displaystyle\sum\limits_{q_i=1...D}(\bar{Y}(q_i, pv_i)-Y(q_i, pv_i))^2 \tag{P3}\label{eq:mse} 
\end{align}

\subsubsection{PID Control System}

In our method, a PID control system is deployed to deal with the changing online environment. We adopt the actuator shown in Eq. \eqref{eq:pid_acuator} in the PID controller, where we regard one hour as a time session. The hyperparameters of $k_p$, $k_i$ and $k_d$ in the PID controller are grid-searched based on the historical data. Especially, we add a multiplier $scaler(t)$ in the actuator since the traffic of the online request in our scenario may change dramatically among hours. We use the $scaler(t)$ as prior knowledge to correct the traffic distribution and improve the feedback control system. The $scaler(t)$ is calculated by the online request number of time session $t$ scaled by the total online request number of the day, which is rather stable in our scenario. In addition, we set the maximum load capability of the system as the reference computation cost (i.e. $C$ in \ref{lp1}) with some tolerable buffer across time sessions to assure online safety. 

\begin{equation} \label{eq:pid_acuator}
\begin{aligned}
x(t+1) = x(0) \cdot exp(-u(t)) \cdot scaler(t)
\end{aligned}
\end{equation}

\subsection{Experimental Results}

In this section, we firstly conduct experiments to illustrate that replacing the original revenue function with logarithm functions results in trivial deviation, and then demonstrate the control capability of the feedback control system. Afterward, we compare our method with the baseline methods on the real dataset in the offline environment. Finally, we deploy our method in the display advertising system of Alibaba, and evaluate its effectiveness in the industrial online environment.

\subsubsection{Revenue Function Fitting Error}

In section 2.2, we propose to approximate the original revenue function by logarithm functions to facilitate the theoretical analysis, and we show the deviation caused by such approximation in this experiment. Although Fig. \ref{revenue_function} gives us the graphical illustration of how neglectable the deviation is, we still need to quantify such deviation. In our evaluation, we show the result in the fine-ranking stage, and other stages deliver similar performance. As stated in problem \eqref{eq:mse}, we aim to minimize the MSE to achieve the approximation. MSE is a good loss function to do optimization, however, it is not an intuitive metric for evaluation since its value changes non-linearly along with the scale of the data. Therefore, we use the main metrics that is commonly adopted in the industrial application, instead of MSE to evaluate the deviation. As shown in Table \ref{tb:revenue_function_fitting}, our evaluation metrics include Mean Absolute Error (MAE),  Mean Absolute Percentage Error (MAPE), Weighted Mean Absolute Percentage Error (WMAPE) and R-Squared Error (R2). These are widely used metrics for approximation and regression problem, and we leave the detailed description of these metrics to the reference \cite{botchkarev2018performance, moksony1990small}. Taking the MAE for example, the average absolute error is $148.85$, which is trivial compared with the average revenue of $3501.80$. In addition, the value of MAPE and WMAPE shows that the deviation compared with the data scale is rather small, which is less than $10\%$. Furthermore, the value of R2 is very close to $1.0$, which means little deviation caused by the approximation. To sum up, we claim that we could replace the original revenue function with the logarithm function to facilitate the theoretical analysis with little influence.

\begin{table}
\begin{center}
\resizebox{1.0\linewidth}{!}{
\begin{tabular}{|c|c|c|c|c|c|c|}
\hline
MAE & MAPE(\%) &  WMAPE(\%) & R2 & Average Revenue   \\ \hline
148.85 & 9.33 & 4.25 & 0.99  & 3501.80 \\ \hline
\end{tabular}
}
\end{center}
\caption{Revenue function fitting errors}\label{tb:revenue_function_fitting}
\end{table}

\subsubsection{Control Capability}

\begin{figure}
 	\centering    
    \includegraphics[width=0.42\textwidth]{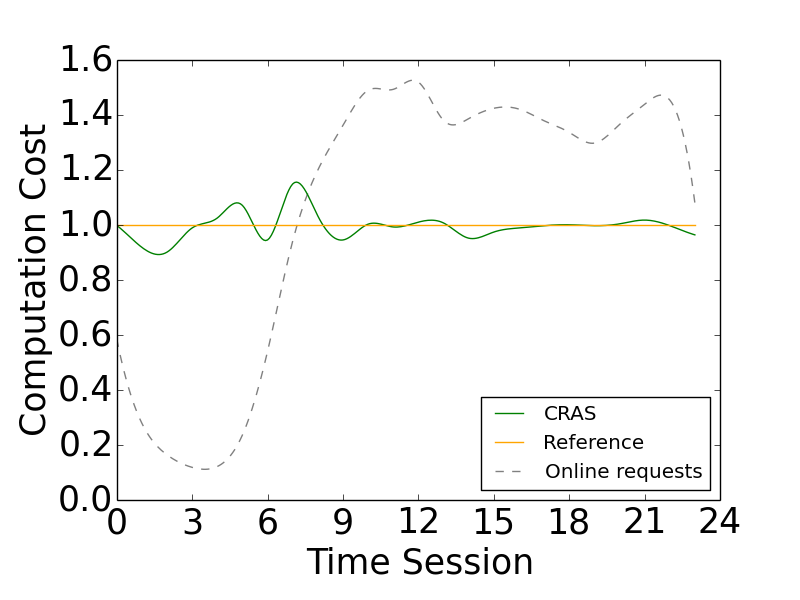}
    \caption{Control capability of the feedback control system. The computation cost is well controlled around $C$.}
    \label{fg:pid_result}
\end{figure}

We conduct this experiment to demonstrate the control capability of the feedback control system. In this experiment, we deploy the feedback control system to adjust the hyperparameter $\alpha$ in the computation resource allocation strategy across continuous time sessions. Please recall that increasing $\alpha$ results in more computation cost. For your information, we conduct this experiment in the fine-ranking stage, and other stages deliver similar performance. As discussed in Section 3.3, we set the constraint $C$ as a reference to control the total computation cost of each time session around it. We illustrate the total computation cost across successive time sessions in Fig. \ref{fg:pid_result}, where $\alpha$ is continuously adjusted by the feedback control system. The horizontal axis is the time session of the day, and the vertical axis is the computation cost. The green line is the computation cost of our method (CRAS), which is continuously controlled by the feedback control system, and the yellow line is the reference computation cost $C$ we want to achieve. In addition, we illustrate the quantity of the online requests in each time session with the dashed line, which demonstrates the significant change of the online environment. As shown in Fig. \ref{fg:pid_result}, the computation cost of our method is well controlled within the margin of the constraint $C$, even with huge changes of online requests. The results show that the feedback control system is able to control the computation cost near the constraint $C$, and thus helps to approach the optimal $\alpha$ of the computation resource strategy in the dynamic online environment.

\subsubsection{Offline Results}

We conduct this experiment to evaluate the effectiveness of our method in each stage independently. In this experiment, we show the performance of the computation resource allocation solution in the coarse-ranking stage and fine-ranking stage respectively, where the response time limit $D$ in each stage is manually set the same as that of the current online system. It is worth noting that we only evaluate the effectiveness of our method independently in each stage in the offline experiments, since the factors that affect the response time across stages such as network transmission is hard to be simulated in the offline environment, which makes the joint effect in the offline environment unreliable. We would evaluate the joint effect across stages with our method in the following online evaluations.  

In the offline evaluation, we compare our method with the baseline method. The baseline method allocates a fixed candidate set size in each stage across online requests, which is widely adopted in industrial practice. Specifically, the baseline method pre-sets the candidate set size for the pre-ranking stage, coarse-ranking stage and fine-ranking stage respectively, and every online request would go through the same truncating process. When we conduct experiments in one specific stage, we keep the candidate set size of other stages equal in the baseline and our method.  

\begin{figure}
 	\centering    
 	\begin{subfigure}[b]{0.23\textwidth}
        \includegraphics[width=\textwidth]{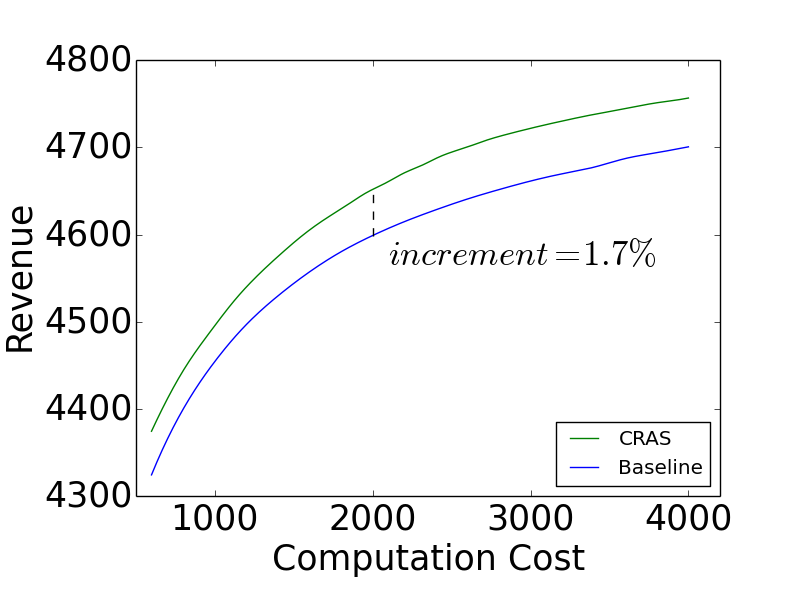}
        \caption{Coarse-ranking stage}

        \label{fg:offline_result_1}
    \end{subfigure}
    \begin{subfigure}[b]{0.23\textwidth}
        \includegraphics[width=\textwidth]{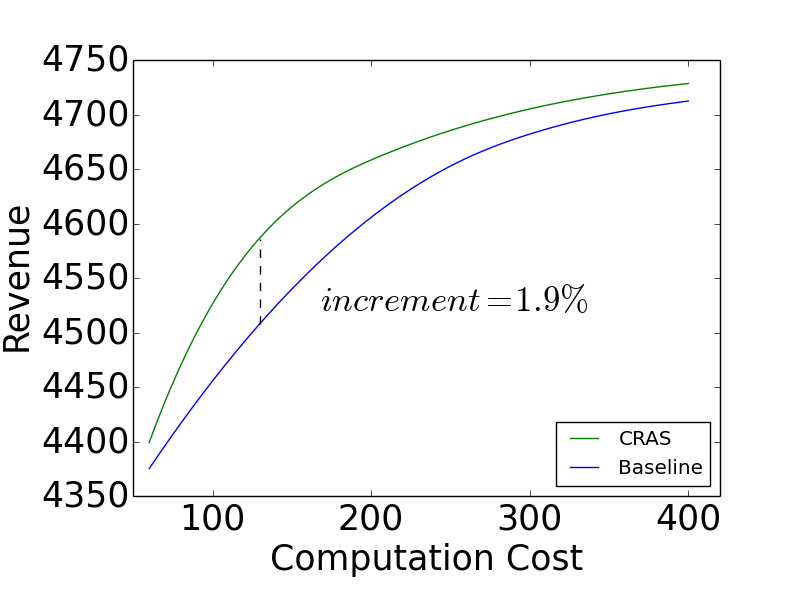}
        \caption{Fine-ranking stage}

        \label{fg:offline_result_2}
    \end{subfigure}
    \caption{Offline results}\label{fg:offline_result}
\end{figure}

We illustrate the offline results in Fig. \ref{fg:offline_result}, where the horizontal axis is the computation cost and the vertical axis is the corresponding revenues. It is worth noting that we use the candidate set size per online request to quantify the total computation cost. We could adjust the fixed candidate set size in the baseline method, and adjust $\alpha$ in our method to control the computation cost. As illustrated in the results, our method (CRAS) significantly outperforms the baseline method in the coarse-ranking stage and fine-ranking stage. As shown in Fig. \ref{fg:offline_result_1} and Fig. \ref{fg:offline_result_2}, our method yields a notable increment of the revenue without increasing any computation cost compared with the baseline method in both stages. We could also compare our method with the baseline method from another perspective. We compare their computation cost with the same revenue, which demonstrates that our method could largely reduce the computation cost without influencing the revenue.

\subsubsection{Online Results}

We deploy our method across stages in the display system of Alibaba and evaluate the joint performance in this experiment. We randomly split the online requests into the buckets of different methods in the online system, and compare their revenues with the same computation cost in the same time session. In addition, we also evaluate the joint performance with different response time allocation in the online experiments. We try different combination of $D_1$, $D_2$ and $D_3$ \footnote{Please refer to InEq. \eqref{rt_constraint1}} in the online experiments to search the optimal response time setting across stages. The summary results are shown in Table \ref {tb:online_results}. We slightly abuse $D_1$, $D_2$ and $D_3$ to represent the fixed candidate set size across stages in the baseline method for better presentation. As demonstrated in the results, our methods (CRAS) yield a significant increment of revenues compared with the baseline method in the industrial online environment. For example, our method improves the revenue by up to $2.60\%$ without increasing any computation cost. Especially, the comparison among our methods with different response time allocation shows that the optimization of the response time allocation could largely improve the business goal in industrial practice. It could be observed in Table  \ref {tb:online_results} that our method $CRAS_4$ with the setting of $D_1 = 12000$, $D_2 = 3500$ and $D_3 = 450$ yields $1.76\%$ more revenues compared with our method $CRAS_1$ with the setting of $D_1 = 10500$, $D_2 = 3500$ and $D_3 = 550$, which demonstrates the efficacy and necessity of our response time allocation framework.
\begin{table}
\begin{center}
\resizebox{1.0\linewidth}{!}{
\begin{tabular}{|c|c|c|c|c|c|}
\hline
  & $D_1$ & $D_2$ & $D_3$ & Revenue & Increment\\ \hline
Baseline &10000& 2000 & 350 & 4356 & 0\% \\ \hline
$CRAS_1$ & 10500 & 3500 & 550 & 4393 & 0.84\% \\ \hline
$CRAS_2$ & 13500 & 2500 & 450 & 4398 & 0.96\% \\ \hline
$CRAS_3$ & 10500 & 4000 & 450 & 4432 & 1.75\% \\ \hline
$CRAS_4$ &12000 & 3500 & 450 & 4469 & 2.60\% \\ \hline
\end{tabular}
}
\end{center}
\caption{Online results}\label{tb:online_results}
\end{table}

\section{Related Work}
	  Online advertising \cite{choi2020online} and recommendation\cite{ davidson2010youtube} are attracting increasing attention in the industry, and many algorithms and strategies have been proposed to improve the business goal of their online systems \cite{wu2018budget, yang2019bid}, where computation cost and response time is not addressed in such work. One general assumption that such previous work holds is that the well-performed models could be applied to the original candidate set of ads, where the cascade-architecture, computation cost and response time constraints in real industrial practice are not considered. As far as we know, this work is the first to maximize the business goal with the consideration of limited computation resources and response time based on the online cascade-architecture. It is worth noting that the framework introduced in this work could be easily combined with previous strategies and algorithms to improve the specific business goal. For example, one could apply certain strategies to maximize a specific business goal, and deploy such strategies across the truncating stages with our method to improve the computation efficiency.

	  As for computation efficiency, there has been quite a lot of work directly addressing the computation efficiency of models. Such work tries to reduce the computation cost of the model by sacrificing minimum estimation accuracy. Most work \cite{hinton2015distilling, han2015deep, he2018amc, zhou2018rocket} achieve computation reduction by simplifying the structure of models. Some work takes advantage of the hardware development \cite{courbariaux2015binaryconnect}, while other work employs the optimization in numerical calculation \cite{gupta2015deep}. The main difference between such work and our work is that such work only considers the computation efficiency of a specific model in a single stage, while our method addresses the computation efficiency with consideration of the joint effect across different models and stages. This recent work \cite{jiang2020dcaf} proposed to allocate computation resources in the granularity of online requests, however, it focuses on one specific stage, where the joint effect across stages and the response time constraint are not addressed.

\section{Conclusion}

In this paper, we propose a computation resource allocation solution that maximizes the business goal of the recommender systems given the computation resources and response time constraints. To the best of our knowledge, this work is the first to address such a problem concerning both computation cost and response time. Specifically, we introduce the common problem that recommender systems are facing, and formulate such a problem as an optimization problem with multiple constraints, which could be broken down into independent sub-problems. Solving the sub-problems, we propose the revenue function to facilitate theoretical analysis and obtain the optimal computation allocation strategy by leveraging the primal-dual method. Especially, the meaning of the optimal strategy could be interpreted from the view of economics. To address the industrial applicability issues, we devise a feedback control system to deal with the changing online environment. Extensive experiments on the real dataset are conducted to demonstrate the superiority of our method. Furthermore, we deploy our method in the display advertising system of Alibaba, and the online results show the effectiveness of our method in real industrial practice.

\balance
\bibliographystyle{ACM-Reference-Format}
\bibliography{acmart}
  
\end{document}